\documentclass[aps,prb,onecolumn,superscriptaddress,floats]{revtex4}
\usepackage{txfonts}
\usepackage{amssymb}
\usepackage{graphicx}
\usepackage{epstopdf}
\usepackage{multirow}
\usepackage{float}
\renewcommand\figurename{\textbf{Figure}}

\begin{document}
\large
\title{Strong Interplay between Stripe Spin Fluctuations, Nematicity and Superconductivity in FeSe
}
\author{Qisi Wang}
\affiliation{
State Key Laboratory of Surface Physics and Department of Physics, Fudan University, Shanghai 200433, China
}
\author{Yao Shen}
\affiliation{
State Key Laboratory of Surface Physics and Department of Physics, Fudan University, Shanghai 200433, China
}
\author{Bingying Pan}
\affiliation{
State Key Laboratory of Surface Physics and Department of Physics, Fudan University, Shanghai 200433, China
}
\author{Yiqing Hao}
\affiliation{
State Key Laboratory of Surface Physics and Department of Physics, Fudan University, Shanghai 200433, China
}
\author{Mingwei Ma}
\affiliation{
Beijing National Laboratory for Condensed Matter Physics, Institute of Physics, Chinese Academy of Science, Beijing 100190, China
}
\author{Fang Zhou}
\affiliation{
Beijing National Laboratory for Condensed Matter Physics, Institute of Physics, Chinese Academy of Science, Beijing 100190, China
}
\author{P. Steffens}
\affiliation{
Institut Laue-Langevin, 71 Avenue des Martyrs, 38042 Grenoble Cedex 9, France
}
\author{K. Schmalzl}
\affiliation{
Juelich Centre for Neutron Science JCNS Forschungszentrum Juelich GmbH, Outstation at ILL, 38042 Grenoble, France
}
\author{T. R. Forrest}
\affiliation{
European Synchrotron Radiation Facility, BP 220, F-38043 Grenoble Cedex, France
}
\author{M. Abdel-Hafiez}
\affiliation{
Center for High Pressure Science and Technology Advanced Research, Shanghai, 201203, China
}
\affiliation{
Faculty of Science, Physics Department, Fayoum University, 63514 Fayoum, Egypt
}
\author{D. A. Chareev}
\affiliation{
Institute of Experimental Mineralogy, Russian Academy of Sciences, 142432 Chernogolovka, Moscow District, Russia
}
\author{A. N. Vasiliev}
\affiliation{
Low Temperature Physics and Superconductivity Department, M.V. Lomonosov Moscow State University, 119991 Moscow, Russia
}
\affiliation{
Theoretical Physics and Applied Mathematics Department, Ural Federal University, 620002 Ekaterinburg, Russia
}
\affiliation{
National University of Science and Technology ``MISiS'', Moscow 119049, Russia
}
\author{P. Bourges}
\affiliation{
Laboratoire Leon Brillouin, CEA-CNRS, CEA-Saclay, 91191 Gif sur Yvette, France
}
\author{Y. Sidis}
\affiliation{
Laboratoire Leon Brillouin, CEA-CNRS, CEA-Saclay, 91191 Gif sur Yvette, France
}
\author{Huibo Cao}
\affiliation{
Neutron Scattering Science Division, Oak Ridge National Laboratory, Oak Ridge, Tennessee 37831-6393, USA
}
\author{Jun Zhao$^\ast$}
\affiliation{
State Key Laboratory of Surface Physics and Department of Physics, Fudan University, Shanghai 200433, China
}
\affiliation{
Collaborative Innovation Center of Advanced Microstructures, Fudan University, Shanghai 200433, China
}

\begin{abstract}

\end{abstract}
\maketitle
\textbf{Elucidating the microscopic origin of nematic order in iron-based superconducting materials is important because the interactions that drive nematic order may also mediate the Cooper pairing \cite{fernandes}. Nematic order breaks fourfold rotational symmetry in the iron plane, which is believed to be driven by either orbital or spin degrees of freedom \cite{fernandes,fernandes2,fang,xu,kruger}. However, as the nematic phase often develops at a temperature just above or coincides with a stripe magnetic phase transition, experimentally determining the dominant driving force of nematic order is difficult \cite{fernandes,dai}. Here, we use neutron scattering to study structurally the simplest iron-based superconductor FeSe (ref.~\onlinecite{cava}), which displays a nematic (orthorhombic) phase transition at $T_s=90$ K, but does not order antiferromagnetically. Our data reveal substantial stripe spin fluctuations, which are coupled with orthorhombicity and are enhanced abruptly on cooling to below $T_s$. Moreover, a sharp spin resonance develops in the superconducting state, whose energy ($\sim 4$ meV) is consistent with an electron boson coupling mode revealed by scanning tunneling spectroscopy\cite{xue}, thereby suggesting a spin fluctuation-mediated sign-changing pairing symmetry. By normalizing the dynamic susceptibility into absolute units, we show that the magnetic spectral weight in FeSe is comparable to that of the iron arsenides \cite{inosov,zhao2}. Our findings support recent theoretical proposals that both nematicity and superconductivity are driven by spin fluctuations\cite{fernandes,fernandes2,fwang,mazin,si,scalapino}.}

  Most parent compounds of iron-based superconductors exhibit a stripe-type long-range antiferromagnetic (AFM) order which is pre-empted by a nematic order: a correlation of electronic states which breaks rotational, but not translational, symmetry. Superconductivity emerges when the magnetic and nematic order are partially or completely suppressed by chemical doping or by the application of pressure \cite{fernandes,dai}. The stripe AFM order consists of columns of parallel spins along the orthorhombic $b$ direction, together with antiparallel spins along the $a$ direction. Similar to the stripe AFM order, the nematic order also breaks the fourfold rotational symmetry, which is signaled by the tetragonal to orthorhombic structure phase transition and pronounced in-plane anisotropy of electronic and magnetic properties \cite{fernandes,dai,yi,chuang,lu,chu}. It has been proposed that nematicity could be driven either by orbital or spin fluctuations, and that orbital fluctuations tend to lead to a sign-preserving $s^{++}$-wave pairing, while spin fluctuations favor a sign-changing $s^\pm$-wave or $d$-wave pairing \cite{fernandes,fernandes2,dai,fang,xu,kruger,scalapino,onari,qzhang}. However, as orbital and spin degrees of freedom are coupled and could be easily affected by the nearby stripe magnetic order, it remains elusive which of them is the primary driving force of nematicity \cite{fernandes,fernandes2,fang,xu,kruger,scalapino,qzhang}.

  FeSe ($T_c\approx8$ K) has attracted great attention not only because of the simple crystal structure (Fig. 1a), but also because it displays a variety of exotic properties unprecedented for other iron based superconductors. For example, the $T_c$ of FeSe increases to $\sim$$40$ K under pressure \cite{cava2} or by ion/molecule intercalation\cite{guo}. In addition, the $T_c$ of single layer FeSe thin film is as high as $100$ K, which is significantly higher than in other iron based superconductors\cite{ge}. More interestingly, unlike most iron-based materials, the tetragonal to orthorhombic structural transition in bulk FeSe is not followed by a stripe magnetic order \cite{cava}, providing an exciting opportunity to elucidate the microscopic origin of nematicity and its interplay with superconductivity. The absence of stripe magnetic order in FeSe seems to cast doubt on the spin driven nematicity scenario. Moreover, recent nuclear magnetic resonance (NMR) measurements suggested that there were little spin fluctuations above $T_s$ in the tetragonal phase, which was also interpreted as a breakdown of the spin scenario \cite{baek,bohmer}. However, NMR only probes momentum-integrated spin fluctuations at very low energies ($\sim 0.1\mu$eV or lower). The momentum dependence of the higher energy spin fluctuations-especially at the energy scale close to the superconducting gap, which is believed to be more important in driving nematicity and superconductivity \cite{fwang,scalapino}-remains unknown. This issue could be addressed by inelastic neutron scattering measurements that probe spin fluctuations over a wide range of momentum and energy.

Neutron scattering studies on FeSe single crystals have been hampered by the lack of high quality samples with the correct phase. Recently, advances in crystal growth with vapor-transport and floating zone techniques have allowed us to grow FeSe single crystals which are significantly larger than what was previously available\cite{chareev,ma}. The superconducting properties of our sample were characterized by DC magnetic susceptibility and resistivity measurements which show an onset $T_c$ of  $8.7$ K with a transition width of $\sim$$0.3$ K, indicating the high quality of the sample (Figs. 1b, 1c). Clear kinks on magnetic susceptibility and resistivity associated with the tetragonal to orthorhombic structure transition are also observed close to $90$ K.

We first use elastic neutron scattering to study the structural and magnetic ordering properties of our FeSe samples. A broadening of the (4, 0, 0) structural peak is observed below $90$ K, indicative of the structural phase transition from the tetragonal to orthorhombic symmetry (Fig. 2a).  The broadened line shape can be fitted with two Gaussian peaks since the sample has orthogonal twin domains and both ($4$, $0$, $0$) and ($0$, $4$, $0$) peaks are covered by the scan. No significant change of the peak width is observed across $T_c$ within our instrumental resolution. On the other hand, no magnetic Bragg peaks associated with the stripe or double stripe magnetic order are observed (not shown) at temperatures down to $1.5$ K, consistent with previous measurements of powder samples \cite{cava}. Instead, in the inelastic channel, we have observed strong spin fluctuations near ($1$, $0$, $0$), which corresponds to the stripe AFM wavevector of the parent compounds of iron-based superconductors \cite{dai}. To determine the momentum dependence of the spin fluctuations and their interplay with superconductivity, we performed rocking/transverse and radial/longitudinal (the scan directions are perpendicular and along $\bf{Q}$, respectively) $\bf{Q}$-scans below and above $T_c$. As shown in Figs. 2b, 2d, representative $\bf{Q}$-scans at $4$ meV are commensurate near ($1$, $0$, $0$) at $T = 11$ K in both transverse and longitudinal directions with no observable anisotropy. The peak intensity is drastically enhanced below $T_c$, which is reminiscent of a magnetic resonant mode observed in other iron-based superconductors \cite{dai,scalapino,inosov,zhao2,zxu,qiu}. Conversely, the scattering at $2.5$ meV is suppressed upon entering the superconducting state due to the opening of the superconducting spin gap (Fig. 2c). The redistribution of the magnetic spectral weight across $T_c$ indicates that the spin excitations near ($1$, $0$, $0$) are closely related to superconductivity. In order to clarify the detailed momentum structure of the  superconductivity-induced magnetic excitations, we have subtracted the signal of the normal state from that of the superconducting state and plot a 2D contour map interpolated from a series of $\bf{Q}$-scans at $4$ meV  (Fig. 2e). The outcome shows that the spin excitation spectra are very sharp with little anisotropy (within our instrumental accuracy). In addition to the results shown near ($1$, $0$, $0$), we also performed similar measurements in the second magnetic Brillouin zone near ($2$, $1$, $0$) associated with the stripe magnetic structure (Fig. 2e). A similar signal is also observed, but with weaker intensity because of the decreased magnetic form factor. These results unambiguously demonstrate that the scattering that we observe here is pure magnetic fluctuation associated with stripe magnetism rather than phonons as the scattering strength from phonons is related to ($\bf{Q}$$\cdot$$\bf{\xi}$)$^2$, where $\bf{\xi}$ is the polarization vector of the phonon.

Figure 3 summarized the energy dependence of dynamic spin correlation function $S$($\bf{Q}$,$\omega$) at $\bf{Q}$=($1$, $0$, $0$) at different temperatures. The figure confirms that the spectral weight loss in the superconducting spin gap ($<3$ meV) is compensated by a sharp resonance mode at around $4$ meV.  Moreover, the detailed temperature dependence of the scattering at $4$ meV shows an order-parameter-like behavior and is clearly coupled to the occurrence of superconductivity (Fig. 4a). The spin resonance mode has been interpreted either as a spin exciton within the superconducting gap arising from scattering between portions of the Fermi surface where the superconducting gap function has an opposite sign \cite{scalapino} or as a broad hump structure induced by overshoot in the magnetic spectrum above the superconducting gap in a sign-preserving $s^{++}$ pairing state \cite{onari}. The sharp mode that we observed here is consistent with the spin exciton model as the mode energy ($4$ meV) is below the superconducting gap ($2\Delta$$\approx$$5$ meV) (ref.~\onlinecite{kasahara1}), and the energy width ($\sim 1.2$ meV) of the mode is essentially resolution-limited and much sharper than in other iron-based superconductors \cite{dai,scalapino,inosov,zhao2,zxu,qiu}. More interestingly, the resonance energy ($E_r$=$4$meV$\approx$$5.3k_BT_c$) is consistent with the electron boson coupling mode ($\sim$$3.8$ meV) revealed by scanning tunneling spectroscopy \cite{xue}, thereby suggesting strong electron-spin excitations coupling in this system. These results are consistent with a spin fluctuation-mediated sign changing pairing mechanism, but inconsistent with an orbital fluctuations-mediated sign-preserving $s^{++}$-wave pairing mechanism \cite{scalapino,onari}.

Although sharp and commensurate stripe spin fluctuations persist at all temperatures measured, the system remains paramagnetic at low temperature. Theoretically, it has been shown that the magnetic interactions in FeSe are much more frustrated than in iron arsenides and therefore prevent long-range magnetic order \cite{fwang,mazin}. Hence, it is informative to compare the magnetic spectral weight in FeSe with that of iron arsenide superconductors. We have calculated absolute units of the imaginary part of the dynamic susceptibility $\chi''$($\bf{Q}$,$\omega$) by normalizing $S$($\bf{Q}$,$\omega$) for the thermal population factor and the intensity of acoustic phonons (Fig. 3b and Supplementary Information). The outcome reveals that the integrated resonance spectral weight ($0.00212$ $\mu_B^2$/Fe) is about $30\%$ of that in the carrier doped BaFe$_{1.85}$Co$_{0.15}$As$_2$ ($E_r=9.5$ meV) (ref.~\onlinecite{inosov}), but two times larger than the damped resonance mode in the isovalently doped BaFe$_{1.85}$Ru$_{0.15}$As$_2$ ($T_c$=$14$ K, $E_r=5.5$ meV) (ref.~\onlinecite{zhao2}). Since the $T_c$ (8.7 K) of FeSe is also about a factor of three lower than in BaFe$_{1.85}$Co$_{0.15}$As$_2$ ($T_c=25$ K), the overall magnetic spectral weight in both systems should be comparable.

Having established the interplay between the spin fluctuations and superconductivity, we now turn to the impact of nematicity on the spin fluctuations. Previous NMR measurements suggested the absence of spin fluctuations above $T_s$ in the tetragonal phase \cite{baek,bohmer}. In contrast, our neutron scattering measurements show substantial spin fluctuations in the tetragonal phase ($T = 110$ K) (Figs. 3a, 2d). We note that the energy dependence of the dynamical spin correlation function $S$($\bf{Q}$,$\omega$) displays a spin gap-like feature at low energies at $T=110$ K (Fig. 3a), which is confirmed by the featureless $\bf{Q}$-scan at $2.5$ meV (Fig. 2c). These results agree with a theoretically predicted gapped nematic quantum paramagnetic state with low carrier density in FeSe (ref.~\onlinecite{fwang}), which naturally accounts for the absence of low energy spin fluctuations above $T_s$ suggested by NMR measurements \cite{baek,bohmer}. The most striking observation is that the spin fluctuations are enhanced abruptly in the orthorhombic phase at $T=11$ K (Fig. 3a). We note that the increase of the spin fluctuations is more pronounced at low energies. To determine if the increase of the spin fluctuation is indeed associated with the nematic order, we carefully measured the temperature dependence of the scattering at $2.5$ meV, which is the lowest energy that can be measured in our thermal triple axis spectrometer with a reasonable background. Intriguingly, a comparison of  the temperature evolution of the $S$($\bf{Q}$,$\omega$) with the orthorhombicity $\delta(T)=(a-b)/(a+b)$ reveals that the enhancement of the $S$($\bf{Q}$,$\omega$) is clearly coupled to the development of the nematic (orthorhombic) phase (Fig. 4b). These results are consistent with the recent proposals (based on either itinerant or local moment pictures) that the nematic order is driven by spin fluctuations \cite{fernandes,fernandes2,fwang,mazin,si}. In a local moment model that frustrated magnetic interactions drive nematic order in FeSe, once the orthorhombic distortion develops, the effective nearest-neighbor exchange couplings $J_{1x}$ and $J_{1y}$ become non-equal and the frustration is partially released, therefore making the system move toward the stripe ordered phase. As a result the spin fluctuations at the stripe ordering wavevector are enhanced.

 It is interesting to compare the spin fluctuations of FeSe with that of iron selenide superconductors without nematic order. The low energy spin fluctuations in FeTe$_{1-x}$Se${_x}$ and Rb$_x$Fe$_{2-y}$Se$_2$ appear at $\bf{Q}$=($1,-0.3\le\xi\le0.3$) and $\bf{Q}$=($1, \pm0.5$), respectively \cite{zxu,qiu,park}. Different from FeSe, the dynamic spin correlation $S$($\bf{Q}$,$\omega$) of FeTe$_{1-x}$Se${_x}$ displays little temperature dependence from $T_c$ to $300$ K (ref.~\onlinecite{zxu}). Moreover, the spin fluctuations of FeTe$_{1-x}$Se${_x}$ are broad and incommensurate/anisotropic \cite{qiu,zxu}, in contrast to the sharp and commensurate spin fluctuations at the stripe AFM wavevector in FeSe. Therefore, FeSe is closer to the stripe magnetic instability and consequently with a larger spin-spin correlation length. These results further imply that nematicity is driven by stripe spin fluctuations, though superconductivity can be mediated by spin fluctuations either at or away from the stripe AFM wavevector.

 In summary, we have reported evidence of strong coupling between the stripe spin fluctuations, nematicity and superconductivity in single crystalline FeSe. Contrary to earlier NMR measurements \cite{baek,bohmer}, our neutron scattering data reveal substantial commensurate stripe spin fluctuations in the tetragonal phase, which are further enhanced in the nematic phase. Moreover, a resolution-limited sharp spin resonance appears well below the superconducting gap and is coupled with electronic density of states, indicating a spin fluctuations-mediated sign-changing pairing symmetry rather than an orbital fluctuations-mediated sign-preserving $s^{++}$-wave pairing symmetry. These results are in agreement with the theoretical predictions that nematicity and superconductivity are driven by spin fluctuations \cite{scalapino,fwang,mazin,si,fernandes,fernandes2}. We believe that the elucidation of the interplay between spin fluctuations, nematicity and superconductivity will have important implications for the understanding of other exotic properties of FeSe, such as the drastically increased $T_c$ under external pressure or substrate strain \cite{ge,cava2,guo}.

 \textit{Note added:} After we finished this paper, we became aware of a related preprint describing neutron scattering measurements on FeSe powder samples \cite{boothroyd}.

$^{*}$Correspondence and requests for materials should be addressed to J.Z. (zhaoj@fudan.edu.cn).

{\bf Acknowledgements}

We thank D. H. Lee, Q. Si, F. Wang and H. Yao for useful discussions. This work is supported by the National Natural Science Foundation of China (Grant No. 11374059) and the Shanghai Pujiang Scholar Program (Grant No.13PJ1401100). M.M. and F.Z. acknowledge support from National Natural Science Foundation of China (Grant No. 11190020). H.C. received support from the Scientific User Facilities Division, Office of Basic Energy Sciences, U.S. Department of Energy. A.N.V. was supported in part from the Ministry of Education and Science of the Russian Federation in the framework of Increase Competitiveness Program of NUST $\langle$MISiS$\rangle$ (No. §¬2-2014-036). D.A.C. and A.N.V. acknowledge also support of Russian Foundation for Basic Research through Grants 13-02-00174, 14-02-92002, 14-02-92693.


\clearpage

\begin{figure}[h]
\includegraphics[width=16cm]{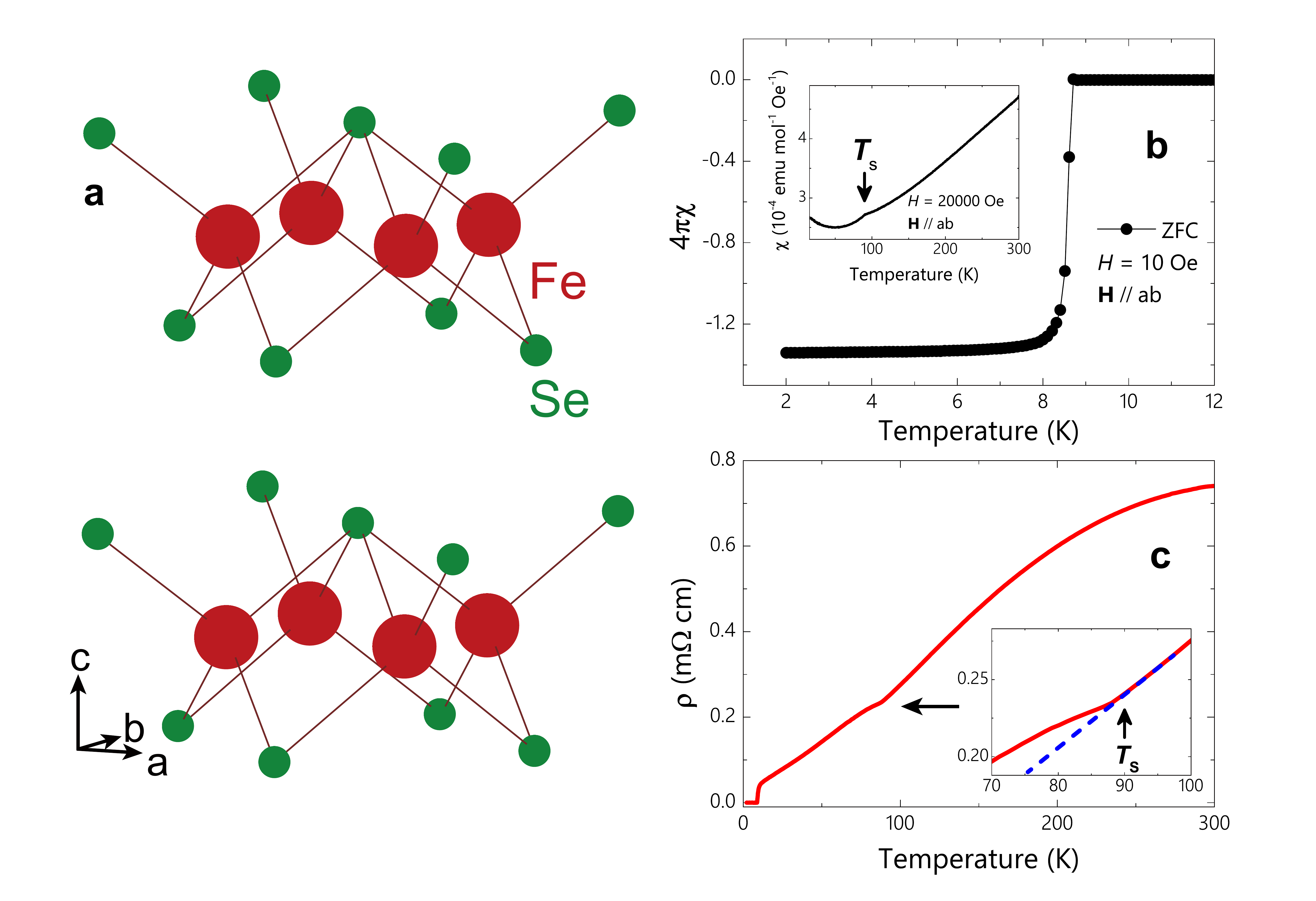}
\caption{ \textbf{Orthorhombic crystal structure, magnetic susceptibility and resistivity of FeSe single crystal}. \textbf{a}, Schematic diagram of FeSe crystal structure. \textbf{b}, The DC magnetic susceptibility measurements on the single-crystalline FeSe sample. A sharp superconducting transition is observed at $T_c = 8.7$ K in the ZFC measurement in a magnetic field of $H = 10$ Oe, indicating $\sim 100\%$ exclusion of the external magnetic field. The screening is slightly larger than $-1$ because of the demagnetization effect. The inset shows the susceptibility measured in a magnetic field of $H = 20$ kOe. The magnetic fields are applied perpendicular to the $c$ axis. \textbf{c}, In-plane resistivity as a function of temperature. The inset shows data around $T_s$ = $90$ K on an enlarged scale.
}

\end{figure}
 \begin{figure*}[h]
\includegraphics[width=16cm]{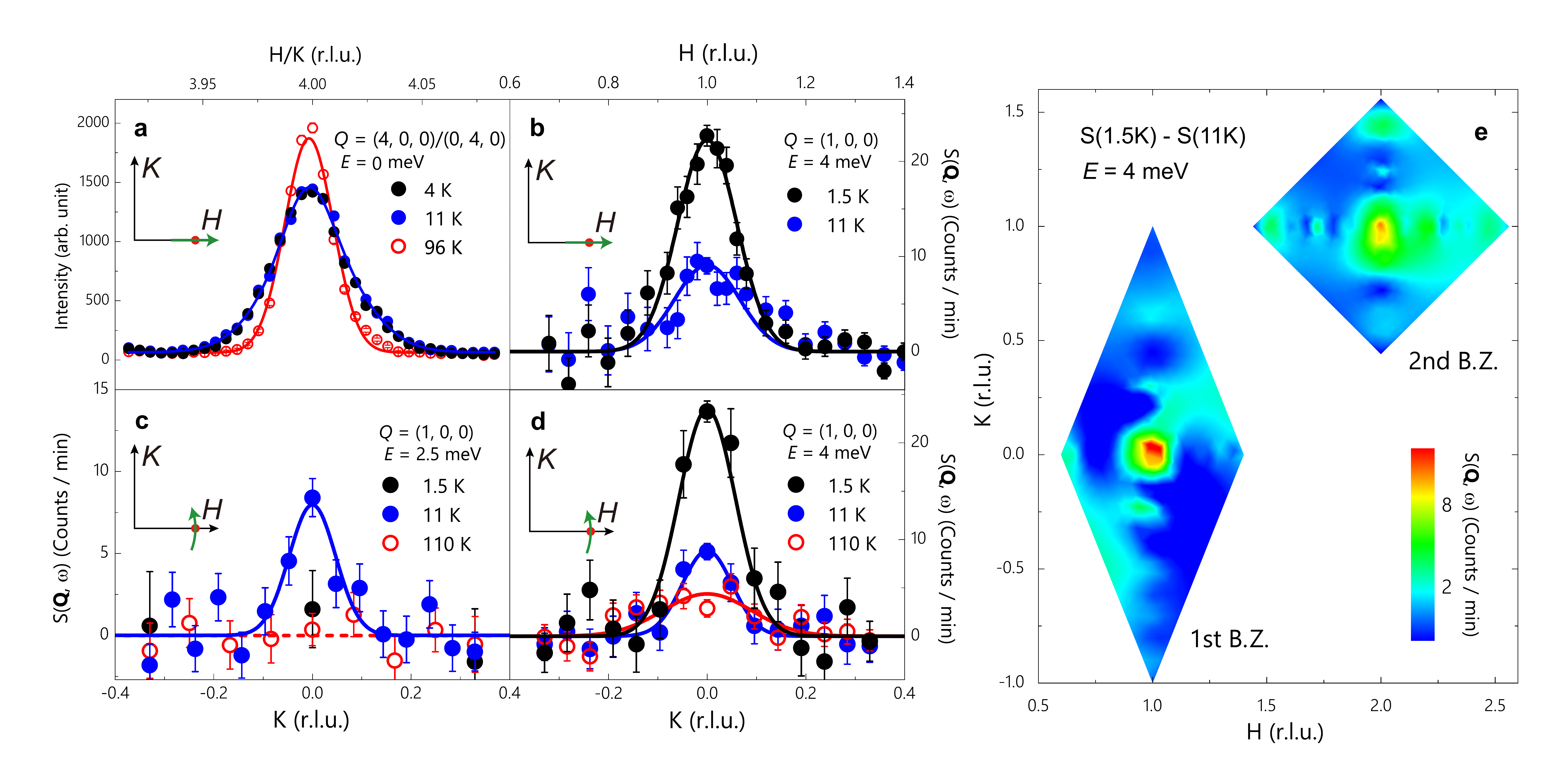}
\caption{\textbf{Structure phase transition and momentum dependence of the spin fluctuations at various temperatures in FeSe}. The inelastic neutron scattering measurements were carried out on the IN20 thermal triple axis spectrometer at the Institut Laue-Langevin, Grenoble, France and the 2T1 thermal triple axis spectrometer at the Laboratoire Leon Brillouin, France. The FeSe single crystals are co-aligned in the (H, K, 0) horizontal scattering plane within $\sim 3$ degrees mosaicity for the measurements. The elastic measurements were performed on one piece of small single crystal on the HB-3A four-circle single-crystal diffractometer at the High-Flux Isotope Reactor at the Oak Ridge National Laboratory, United States (The instrument configurations are described in the Supplementary Information). We present the data by defining the wave vector $\bf{Q}$ at ($q_x$, $q_y$, $q_z$) as ($h$, $k$, $l$)= ($q_xa/2\pi$, $q_ya/2\pi$, $q_zc/2\pi$) reciprocal lattice units (r.l.u.) in the orthorhombic unit cell. \textbf{a}, Temperature dependence of the ($4$, $0$, $0$)/($0$, $4$, $0$) nuclear reflections. The Bragg peak is significantly broadened below the tetragonal-to-orthorhombic phase transition. \textbf{b-d}, $\bf{Q}$-scans near ($1$, $0$, $0$) at various energies and temperatures; linear backgrounds are subtracted (see the Supplementary Information). The scan directions are marked by green arrows in the insets. \textbf{e}, 2D contour plot of the temperature difference scattering [$S(1.5 K)-S(11 K)$] interpolated from a series of $\bf{Q}$-scans at $4$ meV. The error bars indicate one standard deviation.
 }
\label{cha}

\end{figure*}

\begin{figure}[h]
\includegraphics[width=0.6\textwidth]{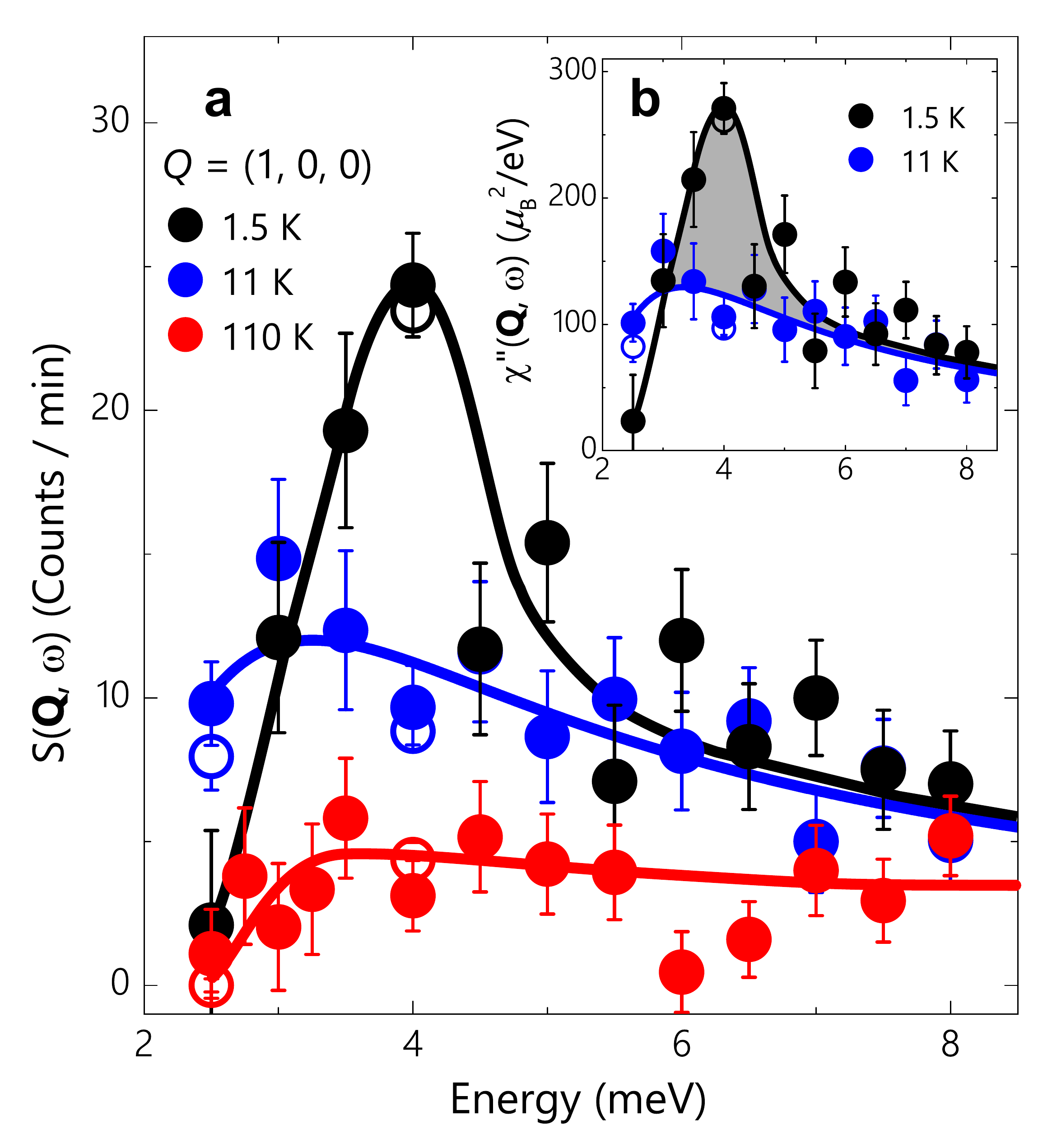}
\caption{\textbf{Energy dependence of spin excitations for FeSe in the superconducting state ($T= 1.5$ K) and normal state ($T = 11$ and $110$ K) } \textbf{a}, Energy dependence of the dynamic spin correlation function $S$($\bf{Q}$,$\omega$) at $\bf{Q}$ =($1$, $0$, $0$) after a background correction. The background is measured at $\bf{Q}$ = ($0.944$, $0.330$, 0) and $\bf{Q}$ = ($0.944$, $-0.330$, $0$), $0$) (see the Supplementary Information). The open circles are data fitted with $\bf{Q}$-scans.  \textbf{b}, Energy dependence of the imaginary part of the dynamic susceptibility $\chi''$($\bf{Q}$,$\omega$) in the superconducting state ($T = 1.5$ K),  and the normal state ($T = 11$ K). The data are obtained from $S$($\bf{Q}$,$\omega$) by correcting for the Bose-population factor and are normalized to absolute units with acoustic phonons as described in the Supplementary Information. The solid curves are guides to the eye. The shaded area denotes the resonance spectral weight. The error bars indicate one standard deviation.
}
\end{figure}

\begin{figure*}[h]
\includegraphics[scale=0.35]{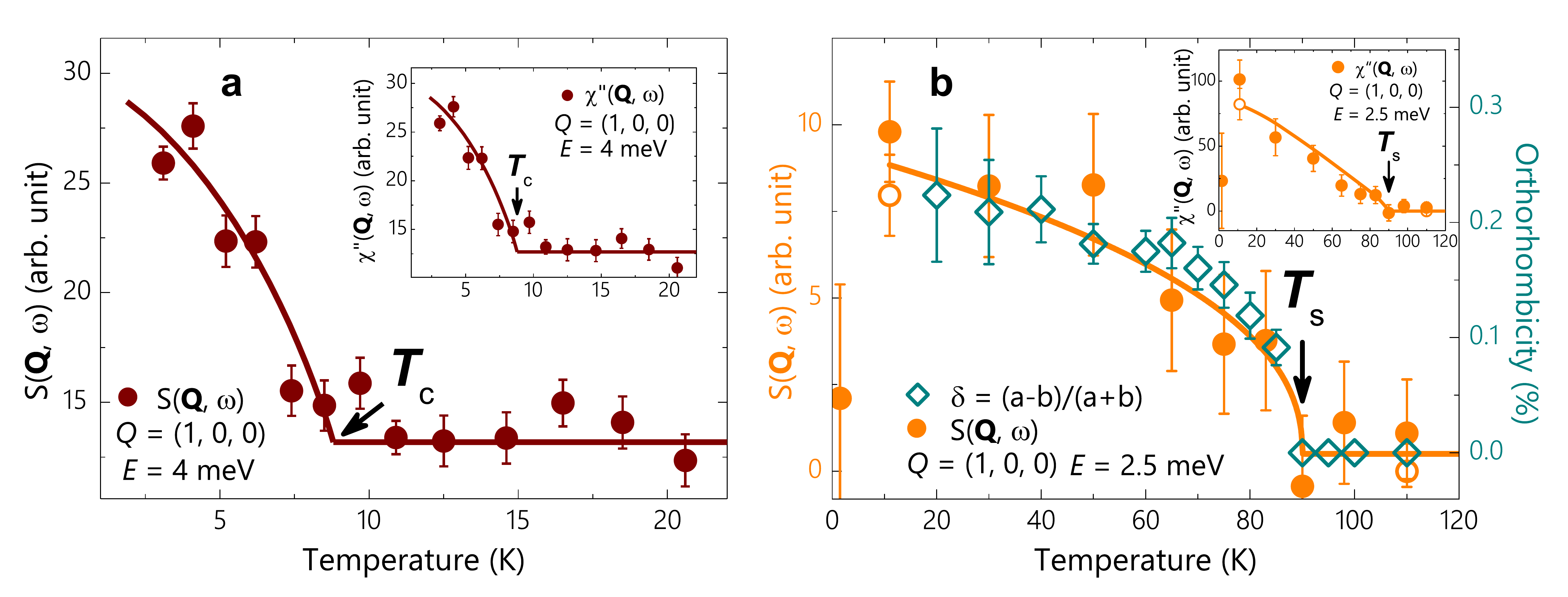}
\caption{\textbf{Temperature dependence of spin fluctuations in FeSe.} \textbf{a}, Temperature dependence of dynamic spin correlation $S$($\bf{Q}$,$\omega$) at $E = 4$ meV, which clearly shows a kink at $T_c$. The inset displays the temperature evolution of $\chi''$($\bf{Q}$,$\omega$). \textbf{b}, Temperature dependence of $S$($\bf{Q}$,$\omega$) at $E = 2.5$ meV and the orthorhombicity $\delta(T)=(a-b)/(a+b)$ shows an order-parameter-like behavior with an onset at $T_s$. The orthorhombicity is adapted from the X-ray diffraction data in ref.~\onlinecite{cava}.  The inset shows the temperature evolution of $\chi''$($\bf{Q}$,$\omega$) which also displays a kink at $T_s$. We note that the decrease of the scattering intensity at $1.5$ K is due to the opening of the superconducting spin gap (Fig. 3). The open circles are data fitted with $\bf{Q}$-scans. The solid lines are a guide to the eye. The error bars indicate one standard deviation.
}
\end{figure*}

\clearpage
\renewcommand\figurename{\textbf{SFig}}
\setcounter{figure}{0}

\maketitle{\textbf{Supplementary Information: Strong Interplay between Stripe Spin Fluctuations, Nematicity and Superconductivity in FeSe}}

\author{Qisi Wang}
\affiliation{
State Key Laboratory of Surface Physics and Department of Physics, Fudan University, Shanghai 200433, China
}
\author{Yao Shen}
\affiliation{
State Key Laboratory of Surface Physics and Department of Physics, Fudan University, Shanghai 200433, China
}
\author{Bingying Pan}
\affiliation{
State Key Laboratory of Surface Physics and Department of Physics, Fudan University, Shanghai 200433, China
}
\author{Yiqing Hao}
\affiliation{
State Key Laboratory of Surface Physics and Department of Physics, Fudan University, Shanghai 200433, China
}
\author{Mingwei Ma}
\affiliation{
Beijing National Laboratory for Condensed Matter Physics, Institute of Physics, Chinese Academy of Science, Beijing 100190, China
}
\author{Fang Zhou}
\affiliation{
Beijing National Laboratory for Condensed Matter Physics, Institute of Physics, Chinese Academy of Science, Beijing 100190, China
}
\author{P. Steffens}
\affiliation{
Institut Laue-Langevin, 71 Avenue des Martyrs, 38042 Grenoble Cedex 9, France
}
\author{K. Schmalzl}
\affiliation{
Juelich Centre for Neutron Science JCNS Forschungszentrum Juelich GmbH, Outstation at ILL, 38042 Grenoble, France
}
\author{T. R. Forrest}
\affiliation{
European Synchrotron Radiation Facility, BP 220, F-38043 Grenoble Cedex, France
}
\author{M. Abdel-Hafiez}
\affiliation{
Center for High Pressure Science and Technology Advanced Research, Shanghai, 201203, China
}
\affiliation{
Faculty of Science, Physics Department, Fayoum University, 63514 Fayoum, Egypt
}
\author{D. A. Chareev}
\affiliation{
Institute of Experimental Mineralogy, Russian Academy of Sciences, 142432 Chernogolovka, Moscow District, Russia
}
\author{A. N. Vasiliev}
\affiliation{
Low Temperature Physics and Superconductivity Department, M.V. Lomonosov Moscow State University, 119991 Moscow, Russia
}
\affiliation{
Theoretical Physics and Applied Mathematics Department, Ural Federal University, 620002 Ekaterinburg, Russia
}
\affiliation{
National University of Science and Technology ``MISiS'', Moscow 119049, Russia
}
\author{P. Bourges}
\affiliation{
Laboratoire Leon Brillouin, CEA-CNRS, CEA-Saclay, 91191 Gif sur Yvette, France
}
\author{Y. Sidis}
\affiliation{
Laboratoire Leon Brillouin, CEA-CNRS, CEA-Saclay, 91191 Gif sur Yvette, France
}
\author{Huibo Cao}
\affiliation{
Neutron Scattering Science Division, Oak Ridge National Laboratory, Oak Ridge, Tennessee 37831-6393, USA
}
\author{Jun Zhao$^\ast$}
\affiliation{
State Key Laboratory of Surface Physics and Department of Physics, Fudan University, Shanghai 200433, China
}
\affiliation{
Collaborative Innovation Center of Advanced Microstructures, Fudan University, Shanghai 200433, China
}


\textbf{I. Instrument configurations for the elastic and inelastic neutron scattering experiments}

Elastic neutron diffraction on FeSe single crystals was measured on the HB-3A four-circle diffractometer at the High Flux Isotope Reactor at Oak Ridge National Laboratory, United States. The neutron energy of $34.4$ meV was used from a bent perfect Si-220 monochromator \cite{hb3a}. We used the high resolution mode by flatting the monochromator (mbend=50) in order to track the lattice distortions (Fig. 2a). The inelastic neutron scattering measurements were carried out on the IN20 thermal triple axis spectrometer at the Institut Laue-Langevin, Grenoble, France, and the 2T1 thermal triple axis spectrometer at the Laboratoire Leon Brillouin, France. For the measurements performed on IN20 (Figs. 2b-2e, 3a, 3b, and 4b), we used a focusing Si(111) as monochromator and a pyrolytic graphite [PG(002)] as analyzer. This setup yielded an energy resolution of about 1 meV at (1, 0, 0) at $E=0$ meV. For the measurements performed on 2T1 (Fig. 3a), PG(002) was used as the monochromator and analyzer. A PG filter was installed in front of the analyzer to eliminate the contamination from the higher-order neutrons. A correction was also made for monitor contamination by higher-order neutrons. For both triple axis spectrometers, the final neutron energy was fixed at $E_f$=$14.7$ meV and no collimation was used.

\textbf{II. Raw data, background subtraction and absolute units normalization}

For typical inelastic neutron scattering experiments, the background is momentum, energy and temperature dependent. SFig. 1 shows several representative raw $\bf{Q}$-scans measured at various temperatures. Each $\bf{Q}$-scan can be fitted by a single Gaussian peak on a linear background. The data presented in Fig. 2 are obtained by subtracting the linear background from the raw $\bf{Q}$-scans.

SFig. 2 shows several temperature difference $\bf{Q}$-scans [$S(1.5 K)-S(11 K)$] at $4$ meV near the first Brillouin zone center (1, 0, 0) and  the second Brillouin zone center (2, 1, 0). The 2D contour plot in Fig. 2e was interpolated from a series of such $\bf{Q}$-scans.

SFig. 3 summarizes the raw energy scans at $1.5$ K, $11$ K and $110$ K. The background was estimated as the average intensity at $\bf{Q}$ = ($0.944$, $0.330$, 0) and $\bf{Q}$ = ($0.944$, $-0.330$, $0$). The background-subtracted data are presented in Fig. 3.

SFig. 4a shows the temperature dependence of the scattering at the signal [$\bf{Q}$=(1, 0, 0)] and background positions at $2.5$ meV. As expected, the scattering intensity of the background decreases gradually with decreasing temperature. Nevertheless, on cooling to below $T_s$=$90$ K, a sudden increase of the scattering intensity at the signal is clearly seen (SFig. 4a). Similar behavior is also observed near $T_c=8.7$ K at $4$ meV (SFig. 4b).

The absolute units of the imaginary part of the dynamic susceptibility $\chi''$($\bf{Q}$,$\omega$) (Fig. 3b) were calculated by comparing the intensity of spin fluctuations with that of the acoustic phonons. This approach has been used to normalize the spin fluctuation intensity in several iron based superconductors \cite{zhaos,inosovs} and is intensively discussed in ref.~\onlinecite{xus}.


$^{*}$Correspondence and requests for materials should be addressed to J.Z. (zhaoj@fudan.edu.cn).


\begin{figure}[h]
\includegraphics[width=16cm]{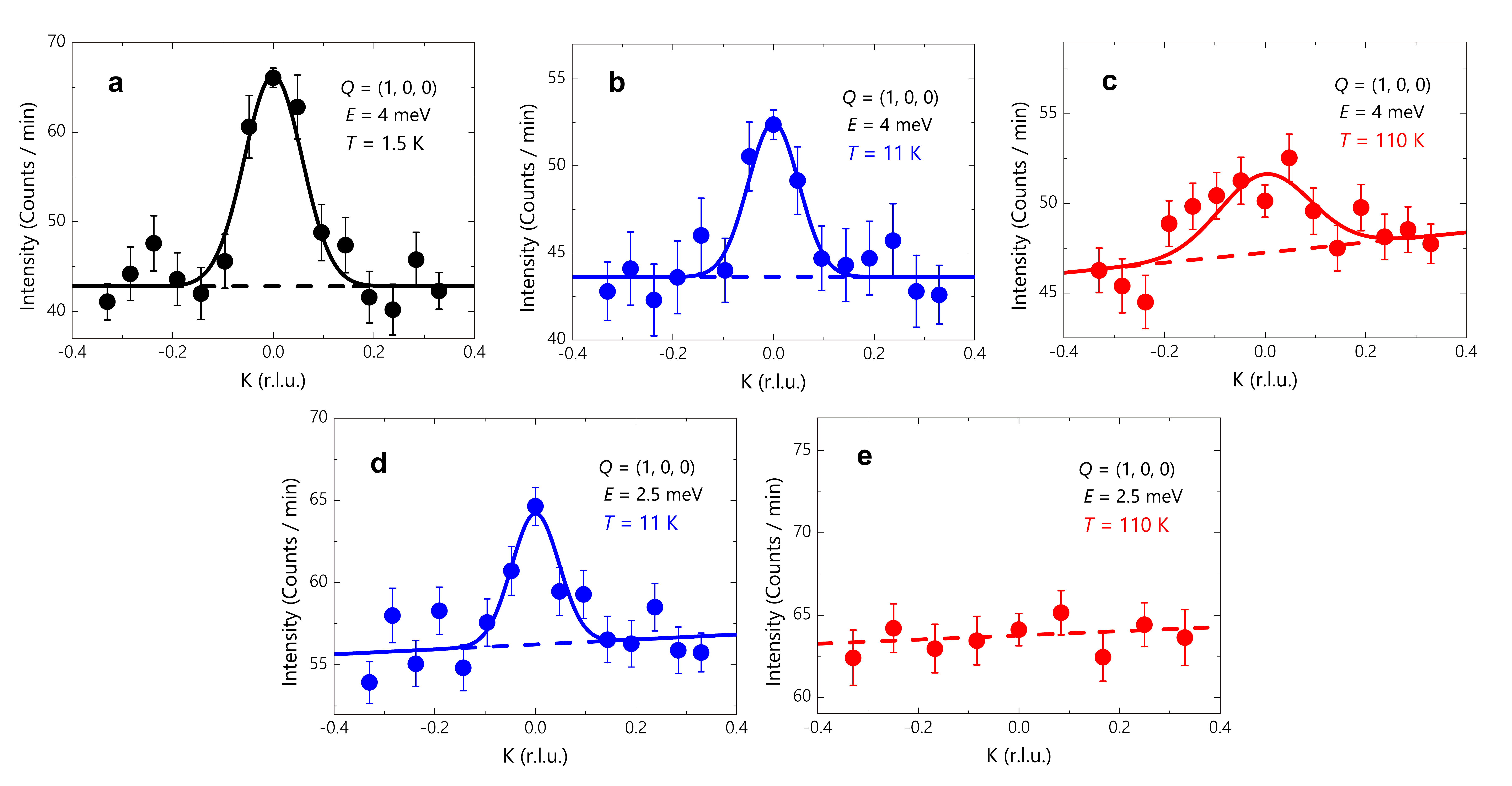}
\caption{ \textbf{Representative raw $\bf{Q}$-scans measured at various temperatures in FeSe}. The $\bf{Q}$-scan can be fitted by a single Gaussian peak on a linear background. The dashed line indicates the background. \textbf{a}, Rocking scan at E=4 meV, Q=(1, 0, 0), T=1.5 K. \textbf{b}, Rocking scan at E=4 meV, Q=(1, 0, 0), T=11 K. \textbf{c}, Rocking scan at E=4 meV, Q=(1, 0, 0), T=110 K. \textbf{d}, Rocking scan at E=2.5 meV, Q=(1, 0, 0), T=11 K. \textbf{e}, Rocking scan at E=2.5 meV, Q=(1, 0, 0), T=110 K.
}
\end{figure}

\begin{figure}[h]
\includegraphics[width=12cm]{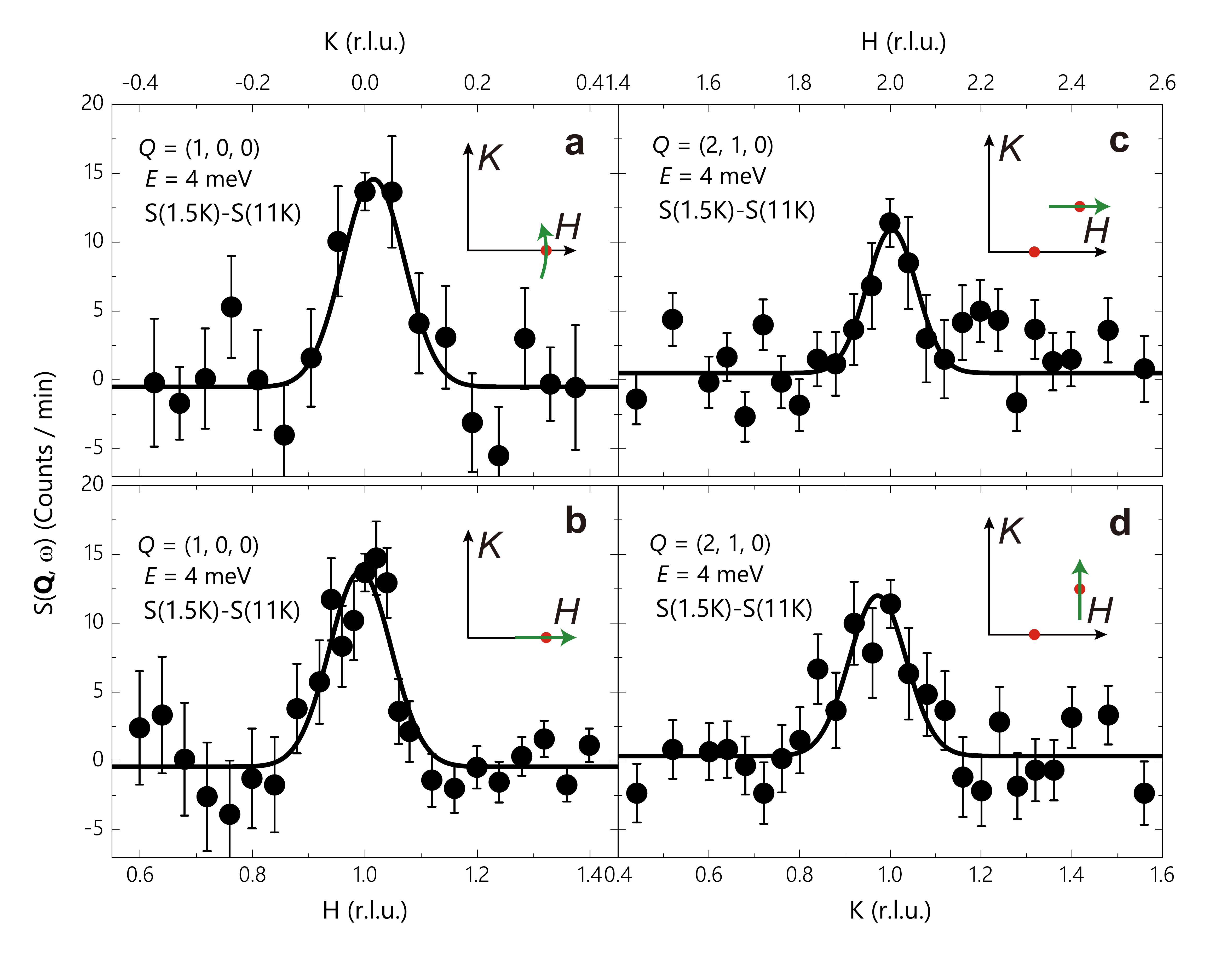}
\caption{ \textbf{Representative temperature difference Q-scans [S(1.5K) - S(11K)] at 4 meV near (1, 0, 0) and (2, 1, 0)}. The 2D contour plot in Fig. 2e was interpolated from a series of such $\bf{Q}$-scans. The scan directions are marked in the insets. Each scan can be fitted by a single Gaussian peak. No significant anisotropy of the peak width is observed. \textbf{a}, Rocking scan at E=4 meV, Q=(1, 0, 0). \textbf{b}, Hscan at E=4 meV, Q=(1, 0, 0). \textbf{c}, Hscan at E=4 meV, Q=(2, 1, 0). \textbf{d}, Kscan at E=4 meV, Q=(2, 1, 0).
}
\end{figure}

\begin{figure}[h]
\includegraphics[width=16cm]{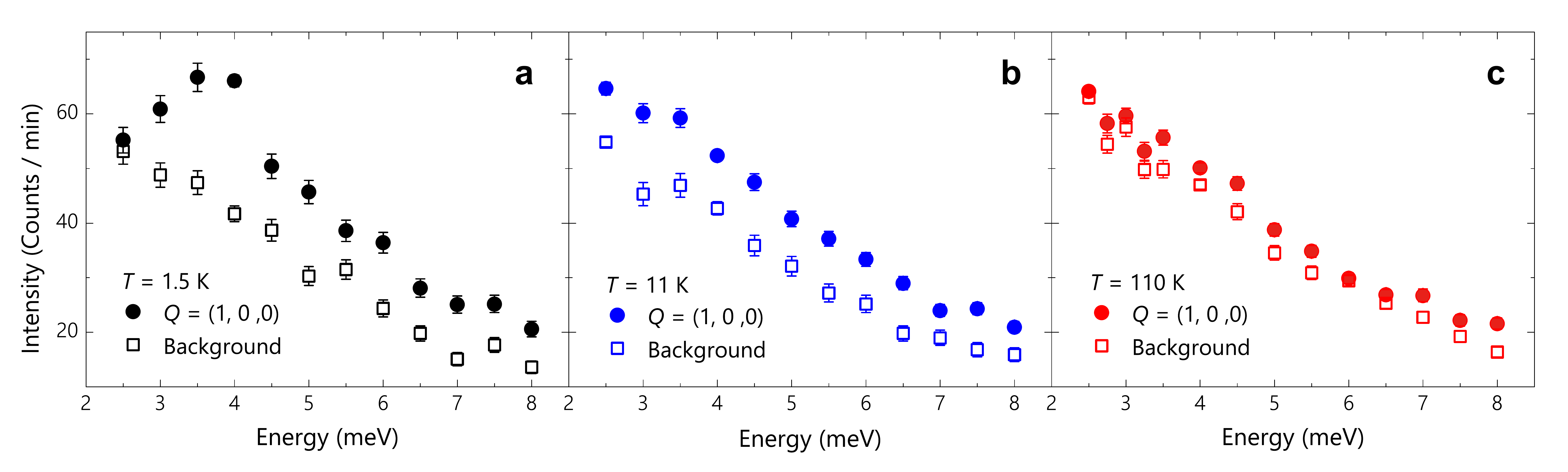}
\caption{ \textbf{Energy dependence of the scattering at the signal [$\bf{Q}$=(1, 0, 0)] and background positions}. The background was estimated as the average intensity at $\bf{Q}$ = ($0.944$, $0.330$, 0) and $\bf{Q}$ = ($0.944$, $-0.330$, $0$). \textbf{a}, T=1.5 K. \textbf{b}, T=11 K. \textbf{c}, T=110 K. The overall magnetic spectral weight is clearly enhanced on cooling from $110$ K to $11$ K at the energies measured.
}
\end{figure}

\begin{figure}[h]
\includegraphics[width=16cm]{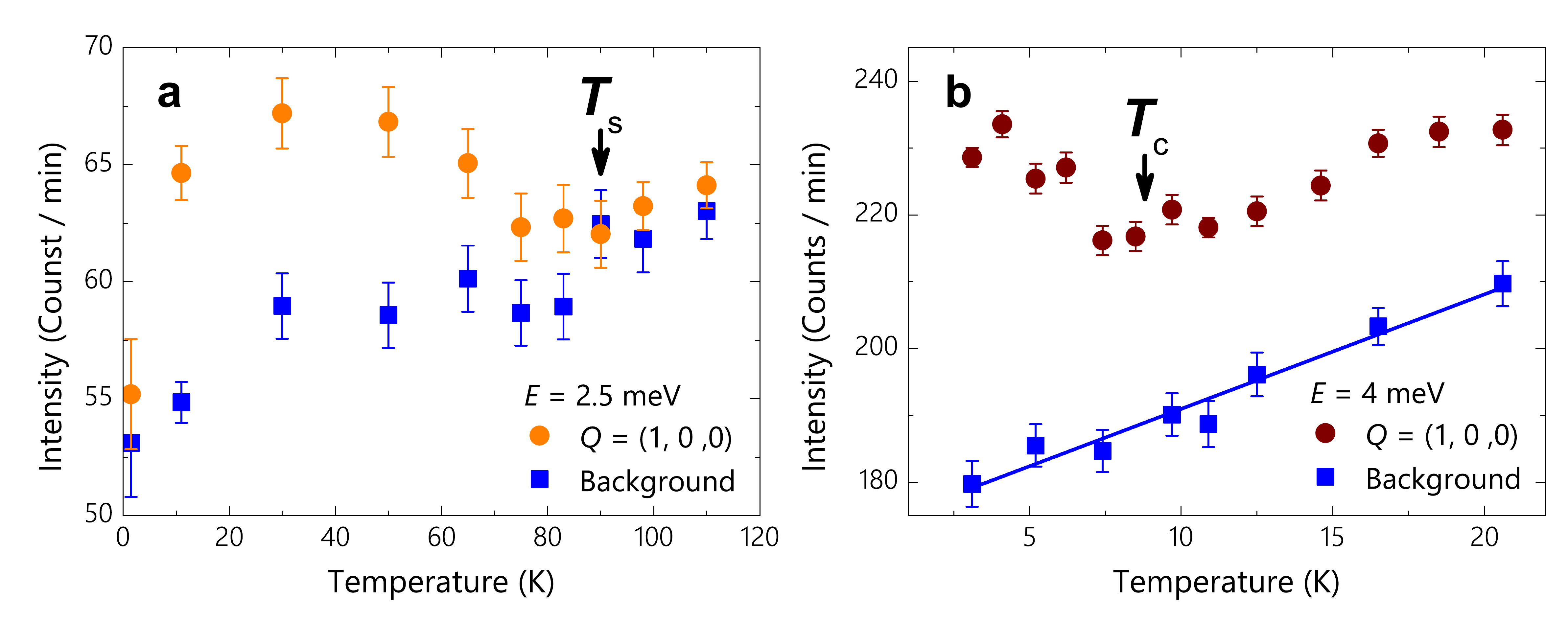}
\caption{ \textbf{Temperature dependence of the scattering at the signal [$\bf{Q}$=(1, 0, 0)] and background positions at $2.5$ meV and $4$ meV}. The background-subtracted data are presented in Fig. 4. \textbf{a}, Temperature dependence of the scattering at the signal [$\bf{Q}$=(1, 0, 0)] and background at $2.5$ meV. The background was estimated as the average intensity at $\bf{Q}$ = ($0.944$, $0.330$, 0) and $\bf{Q}$ = ($0.944$, $-0.330$, $0$). Although the background decreases gradually with decreasing temperature, the signal exhibits a sudden increase at $T_s$=$90$ K. We note that the decrease of the scattering intensity at the signal at $1.5$ K is simply due to the opening of the superconducting spin gap. \textbf{b}, Temperature dependent data for 4 meV with the background measured at $\bf{Q}$=(1, 1, 0) and $\bf{Q}$=(1, -0.6, 0). Since this scan was measured in a relatively narrow temperature range (3K to 21K), the background was estimated by a linear fitting of the data points collected at eight temperatures (blue squares). This is justified as most data points fall on the fitting curve (blue solid line). The data presented in SFig. 4b were collected on 2T1. All other inelastic neutron scattering data were collected on IN20.
}
\end{figure}

\end{document}